\documentclass[a4paper,11pt]{article}
\usepackage{jheppub} 
\usepackage{lineno}
\usepackage{booktabs}
\usepackage{appendix}
\usepackage{color}
\usepackage{float}

\preprint{UT-WI-18-2024}
\title{Plato Meets de Sitter, or de Sitter's Allegory of the Cave}

\author{Willy Fischler}

\author{and Sarah Racz}
\affiliation{Theory Group, Weinberg Institute, Department of Physics,
The University of Texas at Austin, Austin, TX 78712, USA}%

\emailAdd{racz.sarah@utexas.edu}

\abstract{Configurations of masses located at the vertices of Platonic solids deep within the bulk of de Sitter spacetime generate deformations of the cosmological horizon with the geometry dual to these polyhedra. The horizon data encodes both the symmetries and sizes of the solids in the bulk.}

\begin{document}
\maketitle
\flushbottom

\section{Introduction}

For years, physicists have been challenged by the quantum mechanics of gravity. Gravity's holographic nature has emerged as a key concept in this endeavor. For three decades, the holographic principle has been the subject of intense work, in particular in the context of anti-de Sitter spacetime (see \cite{hep-th/9905111} and references therein).  

However, it appears that the universe is accelerating with an equation of state that is consistent with the existence of a small cosmological constant. Therefore, developing a quantum theory of de Sitter spacetime is paramount. Though there has been some preliminary work towards this goal \cite{hep-th/0102077, Fischler2000, Banks2000, hep-th/0007146, hep-th/0609062, hep-th/9806039, 2306.05264,2109.01322,2209.09999} much remains to be done to achieve a satisfactory quantum theory of de Sitter space.

In the 70's Gibbons and Hawking realized that de Sitter space has an entropy \cite{Gibbons1977}. The entropy of de Sitter spacetime is given by the Bekenstein-Hawking area law for the cosmological horizon

\begin{equation}
\label{eq:BekensteinHawking}
S_{dS} = \frac{A_{\mathcal{CH}}}{4 G}.
\end{equation}

Various authors \cite{hep-th/0102077, Fischler2000, Banks2000, hep-th/0007146, hep-th/0609062, hep-th/9806039, 2306.05264,2109.01322,2209.09999,2206.10780} understood from this formulation that empty de Sitter space is maximally entropic, and that any object in the spacetime has the effect of reducing the entropy. In particular, it was shown for the Schwarzschild de Sitter solution (which describes a black hole in de Sitter spacetime) that the presence of that mass reduces the entropy as compared to empty de Sitter. This realization led Banks and Fischler \cite{Fischler2000,Banks2000, hep-th/0007146,hep-th/0102077} and Banks et al \cite{hep-th/0609062} to conjecture that the presence of localized energy-momentum in the bulk corresponds to constraints on the holographic quantum degrees of freedom in holographic space-time (HST), the authors' approach to a holographic theory of de Sitter spacetime. In this context, a natural location for the holographic screen in de Sitter spacetime is the de Sitter horizon \cite{hep-th/0212209}. As we note above, the area of the Schwarzschild de Sitter cosmological horizon has a deficit compared to the area of the de Sitter horizon.
\begin{equation}
    A_{SdS} = A_{dS} - \ell m,
\end{equation}
where $m$ is the mass of the object in the bulk and $\ell$ is the de Sitter radius.

In this paper we will explore more complex configurations of static matter deep within the bulk of de Sitter, exploring their effects on the cosmological horizon. We will show how these non-spherically symmetric configurations of matter deform the de Sitter horizon. We will also show how to extract geometric properties of the objects located deep within the bulk from data on the cosmological horizon. These matter configurations are unstable solutions of Einstein's equations. Any deviation in the location of the masses from their equilibrium position will lead them to either fly towards the horizon in a time proportional to the de Sitter radius or to collapse into a black hole (which quantum mechanically will eventually evaporate), resulting in the empty de Sitter geometry. These configurations are, however, no less stable than the Schwarzschild de Sitter solution.

Specifically, we will use masses with small Schwarzschild radii as compared to the de Sitter radius and arrange them within the bulk, where their gravitational attraction is balanced by the force of cosmic repulsion. A class of static solutions are given by placing the masses at the vertices of Platonic solids. We will show that the masses being small implies that the Platonic solids are also small compared to the radius of de Sitter. The gravitational potential then admits a multipole expansion which we use to examine the deformations of the cosmological horizon generated by configurations of masses located deep within the bulk.

We will show how the symmetries of the Platonic solids are inherited by the horizon, and that, corresponding to each vertex of the solid where a mass is located, the horizon dips compared to the Schwarzschild de Sitter horizon associated to the total mass of the system. Similarly, the faces of the platonic solids correspond to peaks of the aforementioned Schwarzschild de Sitter horizon. In other words, the horizon has the shape of the dual solid to the one in the bulk.

How precisely a quantum theory of de Sitter spacetime accounts for such semi-classical deformations of the horizon associated to such energy configurations in the bulk remains elusive. 

Our paper is organized as follows:
In section 2 we will discuss how the balancing of the gravitational force and the `cosmic repulsion' of masses leads to static albeit unstable configurations deep within the de Sitter bulk. We pay special attention to describing the Platonic solids and their effect on the horizon.

In section 3 we use perturbation theory to incorporate static configurations into the metric. We rely on the formalism originally developed by Regge, Wheeler, and Zerelli to study perturbations of Schwarzschild black holes \cite{Regge1957,Zerilli1970,Guven1990}.

In sections 4 and 5 we use our results from sections 2 and 3 to study how non-spherically symmetric mass affects its shape and present worked through examples.

We conclude with section 6, where we discuss our results and their holographic implications.

\section{Static configurations of multiple masses in de Sitter spacetime}
We begin by finding static configurations of multiple masses within the de Sitter bulk. If the masses are small compared to the de Sitter radius in Planck units (which we adopt for the rest of this paper),  we can find such configurations in the regime where Newtonian gravity is applicable. An arbitrary number of masses may be in a static configuration as long as the gravitational attraction between the masses is balanced by the cosmological expansion. 

Since our focus is the structure of the cosmological horizon in the presence of these small masses, a multipole expansion of the gravitational potential is well suited to describe these configurations. For a mass $m$ separated from the other masses in a given configuration by a distance of size $\mathcal{O}(d)$, the length scale in the multipole expansion, the scale regimes are defined by 
\begin{equation}
\label{eq: dist cond}
    r_{+} \ll d \ll \ell,
\end{equation}
where $r_{+} \equiv 2m$ is the Schwarzschild radius. 

In the most general case, one considers a configuration of $N$ masses, $m_i$ with $i = 1,..., N$. The size of the object within the bulk is fixed by the cancellation of the gravitational attraction between the masses and the cosmological repulsion. The equation of motion for a mass $m_i$ located at position $\mathbf{x}_i$ is given by the Newton-Hooke equation

\begin{equation}
   m_i \frac{\mathrm{d}^2 \mathbf{x}_i}{\mathrm{~d} t^2}=m_i \frac{\mathbf{x}_i }{\ell^2}-\sum_{j \neq i}^{b-N} \frac{m_i m_j\left(\mathbf{x}_i-\mathbf{x}_j\right)}{\left|\mathbf{x}_i-\mathbf{x}_j\right|^3}.
\end{equation}
Static solutions to the above equation satisfy the following constraint 
\begin{equation}
  \frac{\mathbf{x}_i  }{\ell^2}=\sum_{j \neq i}^{b-N} \frac{m_j\left(\mathbf{x}_i-\mathbf{x}_j\right)}{\left|\mathbf{x}_i-\mathbf{x}_j\right|^3}.
     \label{eq:static condition}
\end{equation}
Such static solutions are unstable and maxima of the gravitational potential; they are thus sensitive to perturbations. Small fluctuations will cause the masses either to fly towards the cosmological horizon or to collapse into black hole(s). Arrangements of points satisfying (\ref{eq:static condition}) are called central configurations in the literature and have been studied in \cite{math-ph/0303071,hep-th/0201101,hep-th/0308200} where several physical applications are discussed. 

The center of the mass of the system within the de Sitter bulk must be located at $r=0$. Otherwise the entire matter configuration will fly towards the cosmological horizon. This condition results in there never being an $L=1$ contribution to the multipole expansion of the gravitational potential.

A simple static configuration is that of an equal-mass binary separated by a distance $2d$, which we choose to be aligned along the $z$-axis. Such a configuration of masses in de Sitter spacetime was studied numerically using the Einstein-DeTurck method in \cite{2303.07361}. Solving the static equilibrium condition (\ref{eq:static condition}) for this configuration, we find the location of each mass to be
\begin{equation}
    d_\pm = \pm \left(\frac{\ell^2 m}{4}\right)^{1/3} \mathbf{e}_z.
\end{equation}
Given that in this configuration $m \ll \ell$,  the conditions (\ref{eq: dist cond}) are satisfied, and we conclude that such a configuration of masses exists within the static patch.

Another simple set of solutions to (\ref{eq:static condition}) are given by equal masses located at the vertices of the Platonic solids, a set of 5 convex regular polyhedra; the tetrahedron, the cube, the octahedron, the icosahedron, and the dodecahedron. Each of the platonic solids come with a dual partner for which the faces and vertices of the polyhedron are interchanged. The 5 solids fall into 3 dual pairs according to their symmetry groups, with the tetrahedron being self-dual. We choose to study Platonic solids as they are a particularly simple matter configuration with very little information  required to determine them completely. For Platonic solids centered at the origin, the symmetry groups and Cartesian coordinates of the vertices for the unit solid $(d=1)$ are given below. These facts about the Platonic solids are well known and discussions of the symmetry properties of multipole expansions appear in \cite{Thomson1904,Altmann1957,Gelessus1995}. The coordinate locations of a given Platonic solid are scaled by $d$, which is set by the static condition (\ref{eq:static condition}).

\begin{table}[htbp]
    \centering
    \scriptsize
    \begin{tabular}{lcccc}
        \toprule
        Name  & Symmetry Group & Unit Coordinates & d \\ \midrule
        Tetrahedron  & $A_4$ & $\left(\pm 1, \pm 1, \pm 1\right)$ & $\left( 4 \ell^2 m\right)^{1/3}$\\
        Octahedron  & $S_4$ & $\left(0, 0, \pm 1\right), \left(0, \pm 1, 0\right), \left(\pm 1, 0, 0\right)$ &$\left(\frac{\ell^2 m(1 + 4\sqrt{2})}{4}\right)^{1/3}$\\
        Cube  & $S_4$ & $\left(\pm 1, \pm 1, \pm 1\right)$ & $\left(\frac{\ell^2 m(18 + 9\sqrt{2} + 2 \sqrt{3})}{18}\right)^{1/3}$\\
        Icosahedron  & $A_5$ & $\begin{array}{@{}c@{}}\left(0, \pm 1, \pm \varphi\right), \left(\pm 1, \pm \varphi, 0\right), \\ \left(\pm \varphi, 0, \pm 1\right)\end{array}$ & $\left(\frac{2 \ell^2 m(5 \sqrt{5} + \sqrt{5 - 2 \sqrt{5}})}{5}\right)^{1/3}$\\
        Dodecahedron  & $A_5$ & $\begin{array}{@{}c@{}}\left(\pm 1, \pm 1, \pm 1\right), \left(\pm 0, \pm\frac{1}{\varphi}, \pm \varphi\right), \\ \left(\pm \frac{1}{\varphi}, \pm \varphi, 0\right),\left(\pm \varphi, 0, \pm\frac{1}{\varphi}\right)\end{array}$ & $\left(\frac{\ell^2 m(18 + \sqrt{6(95 + 27\sqrt{10} + 3(\sqrt{6} + \sqrt{15}))})}{36}\right)^{1/3}$\\
        \bottomrule
    \end{tabular}
        \caption{The symmetry group, unit coordinates, and static size is given for every Platonic solid. $\varphi$ is the golden ratio $\varphi = \frac{1+\sqrt{5}}{2} \approx 1.6180$.}
\end{table}

One could construct more complicated distributions of matter within the bulk of de Sitter spacetime involving even more masses which could form interesting polyhedra in the bulk. For example, when $N\geq 13$ the masses cannot lie on a regular polyhedron \cite{hep-th/0201101}. Our methods generalize to such configurations but they will not give qualitatively different results than we show. One could also study mass configurations with multiple length scales. In appendix A.5 we show this by considering two orthogonal pairs of binaries with different masses $m_1$ and $m_2$. 
\section{Perturbations of empty de Sitter spacetime}
We now turn to constructing perturbations of empty de Sitter spacetime that correspond to the static configurations described in the previous section. To accomplish this we employ a straightforward application of Regge-Wheeler formalism, originally developed to study the stability of the Schwarzschild solution. These tools are well suited for studying perturbations about spherically symmetric spacetime. The symmetry of the background allows the perturbations to be expanded into spherical harmonics which decouple from each other. This leads to an infinite set of equations where for each mode $L$ we will solve the perturbation equations. Note that because of the spherical symmetry of the background these equations do not depend on the $M$ (the z-component of the angular momentum). We will associate each perturbation to the multipole expansion of a given configuration's gravitational potential in the weak gravity regime where $d\ll r\ll \ell$.

\subsection{Regge-Wheeler Formalism}
To apply the the Regge-Wheeler formalism  we decompose the spacetime metric $g_{\mu\nu}$ into the background metric $g^{(dS)}_{\mu \nu}$ of empty de Sitter spacetime given by the line element
\begin{equation}
\label{eq: ds Metric}
    ds_{(dS)}^2 = -\left(1-\frac{r^2}{\ell^2}\right)dt^2+\left(1- \frac{r^2}{\ell^2}\right)^{-1}dr^2 + r^2 d\Omega ^2,
\end{equation}
plus a small perturbation $h_{\mu \nu}$. 
After linearizing Einstein's equations for the full metric $g_{\mu \nu}$ we are left with a first order equation for $h_{\mu \nu}$ given by 
\begin{equation} \label{eq:Linear EEQ}
    \delta R_{\mu \nu} - \Lambda h_{\mu \nu} = 0.
\end{equation}
Contained in this expression are both constraint equations, first order in a variable, and dynamical equations, second order in a given variable. Since the constraint and dynamical equations are mixed, the redundancies of Einstein's equations due to the Bianchi identity and the spherical symmetry of the background metric become obscured.

The variation of the Ricci tensor $\delta R_{\mu \nu}$ for the full spacetime is given by the difference of variations of the affine connection, 
\begin{equation}
    \delta R_{\mu \nu} = \delta \Gamma_{\mu \nu ; \beta}^\beta - \delta \Gamma _{\mu \beta ; \nu }^\beta,
\end{equation}
which is given by
\begin{equation}
    \delta \Gamma_{\mu \nu}^\sigma = \frac{1}{2}g^{\sigma \lambda}\left(h_{\mu \lambda ; \nu}+h_{\nu \lambda;\mu}-h_{\mu \nu ; \lambda} \right),
\end{equation}
where the covariant derivative is defined with respect to the background metric. We note that while $\Gamma^\alpha_{\beta\gamma}$ is not a tensor, its variation $\delta\Gamma^\alpha_{\beta\gamma}$ is.

\subsection{Form of perturbations}
Due to the spherical symmetry of the background, the perturbations $h_{\mu \nu}$ can be decomposed into harmonics on the sphere. These spherical harmonics can be further decomposed into components with even and odd parities called polar and axial perturbations, respectively. For our analyses we choose the static polar perturbations in ``Regge-Wheeler gauge" which have the form 

\begin{equation}
\label{eq:polar perturbation}
h^{(L,M)}_{\mu \nu }= Y_{LM}\left(\theta,\phi\right) 
\begin{pmatrix}
H_0^{(L,M)}(r)(1-r^2/\ell^2) & 0 & 0 & 0\\
0 & \frac{H_2^{(L,M)}(r)}{(1-r^2/\ell^2)}  & 0 & 0\\
0 & 0 & r^2 K^{(L,M)}(r) & 0\\
0 & 0 & 0 & r^2 \sin^2\theta K^{(L,M)}(r)
\end{pmatrix}
\end{equation}
where $Y_{LM}$ is the spherical harmonic, and $H_0^{(L,M)}$, $H_2^{(L,M)}$, and $K^{(L,M)}$ are all functions of coordinate $r$ and have implicit dependence on $L$ and $M$. 

Solutions to the linearized Einstein's equations that correspond to matter configurations of the bulk will comprise of superpositions of the $h_{\mu \nu}^{(L,M)}$ to capture the angular dependence of the Newtonian multipole expansion. 

\subsection{Solutions of the perturbation equations}
We now use our explicit form of the perturbation to derive constraints on the functions $H_0^{(L,M)}(r), H_1^{(L,M)}(r)$ and $K^{(L,M)}(r)$ for a given  $(L, M)$. Solving (\ref{eq:Linear EEQ}) leads to 5 non-vanishing components of the linearized Einstein equations, $\delta R_{tt}$, $\delta R_{rr}$, $\delta R_{\theta \theta}$, $\delta R_{\phi \phi}$, and $\delta R_{r\theta}$. These 5 equations lead to three constraints, one algebraic and two differential. The equations to determine $H_0^{(L,M)}, H_2^{(L,M)},$ and $K^{(L,M)}$ are all $M$-independent due to the spherical symmetry background so the radial equations only depend on the mode $L$. The algebraic constraint leads to significant simplifications of the equations and we define a new function $H^{(L)}(r)$, given by
\begin{equation}
H^{(L,M)}(r) = H_2^{(L,M)}(r) \equiv H^{(L,M)}(r).
\end{equation}
The remaining two constraints lead to coupled second order differential equations for the functions $H^{(L,M)}(r)$ and $K^{(L,M)}(r)$ given by
\begin{equation}
\label{eqn:RW 1}
    0= H^{(L,M)}(r) \frac{f'}{f} + \frac{d}{dr} \left(H^{(L,M)}(r)-K^{(L,M)}(r)\right)
\end{equation}

\begin{equation}
\label{eqn:RW 2}
    \begin{aligned}
         0&=H^{(L,M)}(r) \left( \frac{f'}{f} - \frac{f''}{f'}\right) + \frac{d}{dr} H^{(L,M)}(r) \\&+ \frac{(H^{(L,M)}(r)-K^{(L,M)}(r))}{r^2 f'} \left( L\left(L+1\right) -r^2 \Lambda - 2 f - 2 r f' \right)=0,
    \end{aligned}
\end{equation}
where $f = 1-r^2/\ell^2$.
For a given value of $L$ the solutions of these equations give rise to a linear combination of hypergeometric functions. The constants of integration, $\mathbf{c}_{(1)}^{(L,M)}$ and $\mathbf{c}_{(2)}^{(L,M)}$  are fixed by matching to the multipole expansion of the Newtonian potential, such as (\ref{eq:potential cube}) for the cube. For a given value of $L$ we have $4 L + 2$ constants to match (2 for each potential value of $M$ in the superposition). We are however immediately able to set  $2L + 1$ constants to 0 as they do not have the appropriate $r$ behavior. To simplify our expressions we now drop the numerical subscript on our constants of integration. We are left with solutions of the form 
\begin{equation}
\label{eq:H general}
    H^{(L,M)}(r)= \frac{r^L \left(i e^{-i \pi  L} \mathbf{c}^{(L,M)} \left(\frac{\ell}{r}\right)^{2 L+1} \, _2F_1\left(\frac{1}{2} (-L-1),-\frac{L}{2}-1;\frac{1}{2}-L;\frac{r^2}{\ell^2}\right)\right)}{\ell^2-r^2}
\end{equation}

and 

\begin{equation}
\begin{aligned}
K^{(L,M)}(r) &= \frac{-i \ell \mathbf{c}^{(L,M)}}{\left(\ell^4 \left(r^2-\ell^2\right)\right) ((L-1) r)}\left(-\frac{r}{\ell^2}\right)^{-L} \Bigg(\frac{2 (L+1) r^4 \, _2F_1\left(\frac{1-L}{2},-\frac{L}{2};\frac{3}{2}-L;\frac{r^2}{\ell^2}\right)}{2 L-1}\\&+\left(l^4 (L-1)+2 \ell^2 r^2\right) \, _2F_1\left(\frac{1}{2} (-L-1),-\frac{L}{2}-1;\frac{1}{2}-L;\frac{r^2}{\ell^2}\right)\Bigg).
\end{aligned}
\end{equation}

\section{Finding the warped horizon}
With metric in hand, we now turn our attention to finding the cosmological horizon. Since the configurations in the bulk are not spherically symmetric, the horizon will be a null surface $r(\theta,\phi)$ with angular dependence. To find this null surface we use the well-known technique of finding where the one-form $\partial_\mu r$ becomes null or $g^{rr} = 0$, which is valid to linear order in our expansion for the reasons outlined in \cite{arXiv:0707.3222}.  Working to first order in the perturbation $h_{\mu \nu}^{(L,M)}$, the null condition for each is given by
\begin{equation}
\label{eqn:horizon Condition}
    g^{rr} = g^{rr}_{(dS)} - h^{rr} = 1 - \frac{_{\mathcal{H}}r^2}{\ell^2} - (1-\frac{r_{\mathcal{H}}^2}{\ell^2})\sum_L\sum_M H^{(L,M)}(r_{\mathcal{H}}) Y_{LM}(\theta, \phi) = 0,
\end{equation}
where $r_{\mathcal{H}}$ is the location of the cosmological horizon, which in general will depend on $\theta$ and $\phi$. It is understood that the functional form of $H^{(L,M)}(r)$ is the same for every $M$ but the constants may differ.  

\subsection{L=0 Perturbations}
The leading term in the multipole expansion is the $L=0$ mode, which resembles a point-like object of mass $m$ located at $r=0$. The $L=0$ perturbation functions are given by
\begin{equation}
    H^{(0,0)}(r) = \frac{4 \sqrt{\pi} m \left(\ell^2+r^2\right)}{r \left(\ell^2-r^2\right)} \quad \& \quad K^{(0,0)}(r) = \frac{2m}{r}, 
\end{equation}
where the constants of integration were found by matching (\ref{eq:H general}) to the Newtonian monopole potential $\Phi = -m/r$ in the limit $r/\ell \ll 1$. We then substitute $H^{(0,0)}(r)$ into (\ref{eqn:horizon Condition}) to find the new cosmological horizon, which will be spherically symmetric since the $L=0$ mode has no angular dependence. We find the new horizon location to be $r_{\mathcal{H}} = \ell - 2 m$. 

We expect that the $L=0$ mode should correspond to the expansion of the Schwarzschild de Sitter solution to first order in the mass. To compare our results to the linearized Schwarzschild de Sitter solution we need to take care of the $K^{(0,0)}(r)$ perturbation function which appears in front of the angular components of the perturbed metric. We do so by transforming the radial coordinate $r$ into the Schwarzschild de Sitter coordinate $r'$ through the coordinate transform $r' = r \sqrt{\frac{2 m}{r}+1} $. Applying this transformation to the coordinate location of the cosmological horizon we recover the Schwarzschild de Sitter static coordinate horizon location $r'_{\mathcal{H}} = \ell - m$.

\subsection{L$\geq$ 2 Perturbations}
Now that we have shown our methods reproduce the known results of the Schwarzschild-de Sitter spacetime we can turn our attention to $L\geq 2$ perturbations that introduce angular dependence to the horizon. 

For every shape in the bulk we will restrict ourselves to the leading non-vanishing term, $L_{min}$, in the multipole expansion which characterizes the dominant angular contribution to the cosmological horizon. The value of $L_{min}$ will depend on the symmetries of the bulk matter configuration. We build upon the horizon shift due to the $L=0$ perturbation and capture the higher $L$ corrections to the horizon through the following ansatz
\begin{equation}
    r'_{\mathcal{H}} = \ell - m -\epsilon(\theta,\phi),
    \label{eq: horizon ansatz}
\end{equation}
where $r'_{\mathcal{H}}$ is the Schwarzschild de Sitter static coordinate. 
This ansatz is substituted in to (\ref{eqn:horizon Condition}) along with the $H^{(L,M)}$ perturbation functions for $L=0$ and $L=L_{min}$ and solved for $\epsilon(\theta,\phi)$ which characterizes the angular corrections to the horizon. As we will see, the angular dependence inherits the same symmetries as the static mass configuration within the bulk.

\subsection{Area of the horizon}
Having established how the location of the horizon changes for various cases we now turn our attention to the surface area of the cosmological horizon. 

The area element of the induced metric on the horizon to leading order in the metric perturbation is given by
\begin{equation}
    \sqrt{|g|} = r^2 \sin(\theta) \left( 1 + \sum_L\sum_M K^{(L,M)}(r) Y_{LM} (\theta,\phi)\right),
\end{equation}
where we sum over the relevant $L$ and $M$ given a distribution of matter in the bulk. One can immediately see that, apart from the $L=0$ mode, there will be no contribution to the area of the horizon from the perturbations due to the orthogonality of the spherical harmonics. We thus find that to leading order in the perturbations, the only effect that contributes to the change in area of the cosmological horizon is the total mass in the spacetime.

We will now study the area-preserving effects on the cosmological horizon of these higher modes. For any perturbation with $L\geq 2$, the horizon dips and protrudes in such a way that the net effect on the area cancels out. What we will show is that the surface area of one of these dips (or protrusions) encodes the size of the object deep within the bulk. The simplest feature to study will be located at the north pole of the horizon at $\theta=0$ and extends to an angle $\theta_{crit}$,where the horizon size is given by correction due to the $L=0$ perturbation. In other words, between the dips and protrusions, the horizon size is that of the leading Schwarzschild de Sitter solution of the total mass in the spacetime.

We find the surface area by integrating over a spherical cap where $\phi$ runs from 0 to $2\pi$ and $\theta$ goes from 0 to $\theta_{crit}$. This subregion of the cosmological horizon is gauge independent. Further, that partial surface area that is manifestly dependent on $d$, the parameter which gives the size of the object within the bulk. We have thus been able to show that not only the shape of the bulk matter distribution, but also its size, is encoded in the cosmological horizon.

\section{Examples}
We now present some worked examples of how non-spherically symmetric matter configurations in the bulk affect the cosmological horizon. The procedure for finding the warped horizon is first to calculate the multipole expansion of the Newtonian potential $\Phi$ for the given configuration of matter. Next, we match the constants of integration on the perturbation functions $H^{(L,M)}$ and $K^{(L,M)}$ to $\Phi$ in a region of space where $d\ll r \ll \ell$. Finally, we use the perturbed metric to characterize the horizon.

\subsection{Static mass binary}
For the pair of static masses aligned along the $z$-axis the multipole expansion of the Newtonian potential to leading order in $d$ is given by
\begin{equation}
\label{eq:z potential}
   \Phi_{dipole}=-\frac{m}{r} -\frac{d^2 m (3 \cos (2 \theta )+1)}{4 r^3},
\end{equation}
where $d$ is given by (\ref{eq: z location}). Having already considered the $L=0$ perturbation in section 4, we study the next lowest order term in the multipole expansion, $L=2$. Since the masses are aligned in the $z$-direction, the angular dependence is given by the spherical harmonic $Y_2^0$. Thus, we are able to set all of the constants of integration aside from $\mathbf{c}_2^0$ to 0. After matching the coefficient of $H^{(2,0)}(r)Y_2^0(\theta,\phi)$, given in (\ref{eq:H general}), to the $L=2$ term of the Newtonian potential, we find 
\begin{equation}
     H^{(2,0)}(r)Y_2^0(\theta,\phi) = \frac{d^2 m \left(3 \cos ^2(\theta )-1\right) \left(\ell^2-r^2\right) }{\ell^2 r^3}
\end{equation}
and
\begin{equation}
K^{(2,0)}(r) Y_2^0(\theta, \phi) = \frac{ d^2 m \left(3 \cos ^2(\theta )-1\right) \left(\ell^2+r^2\right)}{\ell^2 r^3}.
\end{equation}
after substituting in these expressions into the null surface condition.

Using our horizon location ansatz and this metric perturbation, we find the coordinate location of the cosmological horizon which is given by 
\begin{equation}
    \label{eq: z location}
    r'_{\mathcal{H}} = \ell - m + d^2 m^3 \left(-\frac{12 \cos (2 \theta )}{\ell^4}-\frac{4}{\ell^4}\right),
\end{equation}
\text{after being transformed to a coordinate that allows us to compare to linearized SdS.} 
From this equation, we see that the first order contribution for $L=2$ is small as expected, however, it is the leading non-spherical correction to the horizon consistent with the angular extension of the object in the bulk. Note that the $L=0$ component may get higher order corrections that might dominate this term, but obviously do not affect the angular dependence. We plot (\ref{eq: z location}), for a value of the total mass $m$ outside the range of perturbative validity to qualitatively show the behavior of the horizon. We plot the surfaces using a temperature map where larger values of the coordinate $r'$ are shown in red and smaller values in blue. In the figure below, we see that the sphere is squashed along the $z$-axis along the direction of the masses in the bulk. This aligns with intuition from the Schwarzschild de Sitter spacetime that masses cause the horizon to shrink in that the horizon dips further towards the localized mass-energy. 
\begin{figure}[h]
    \centering
    \includegraphics[width=8cm]{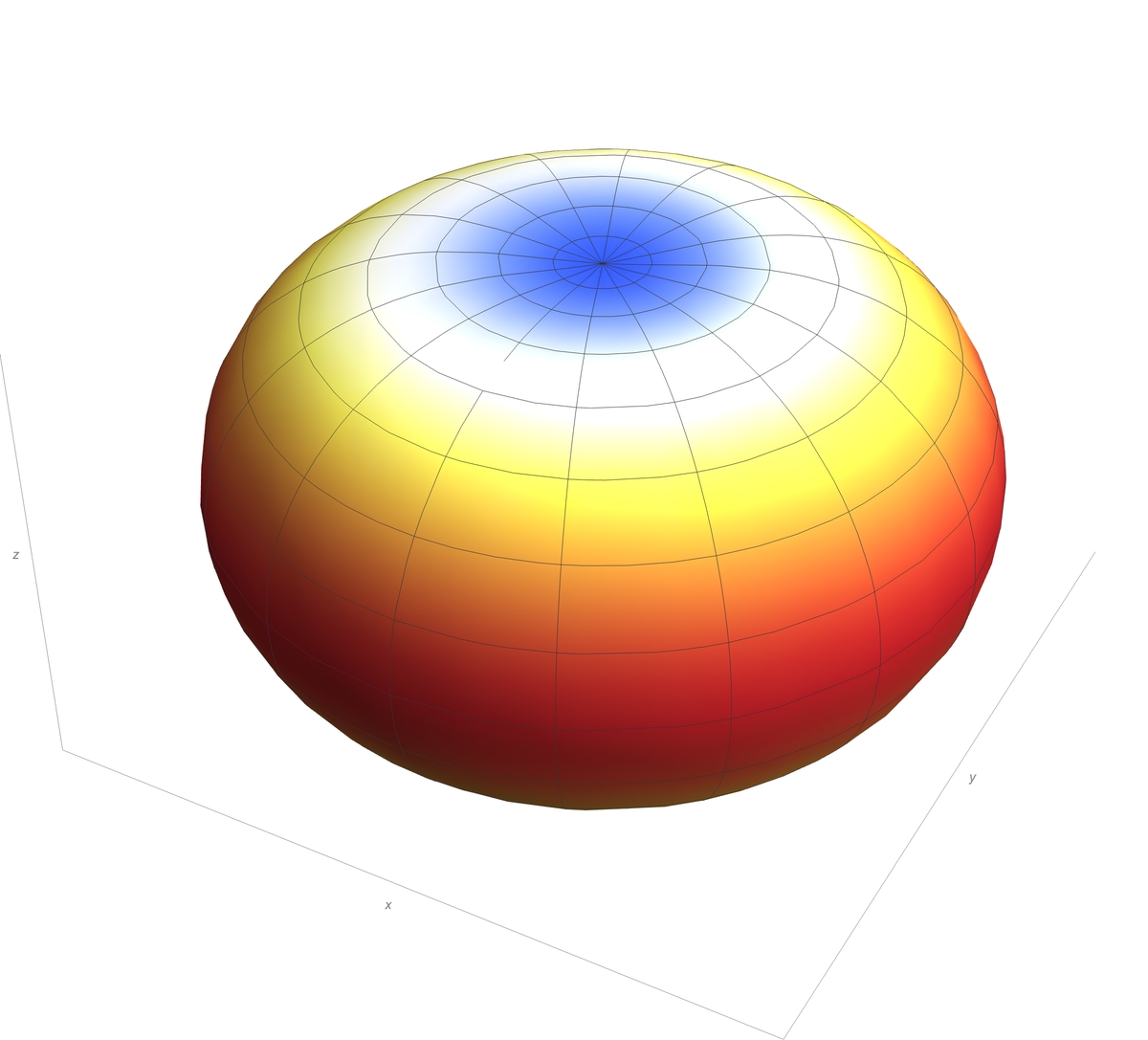}
    \caption{The cosmological horizon for a pair of static masses in the bulk plotted in static coordinates. The horizon is shown for an extreme value of $m$, beyond perturbative validity, to highlight the qualitative features of the horizon's shape. The coloring of the horizon is a temperature map with red and blue corresponding to the largest and smallest values of $r'$ respectively. The plot shows that the horizon moves inwards towards the masses leading to a squashed sphere geometry. }
    \label{fig:zdipole}
\end{figure}

\subsection{Cube}

\begin{figure}[H]
    \centering
    \includegraphics[width=8cm]{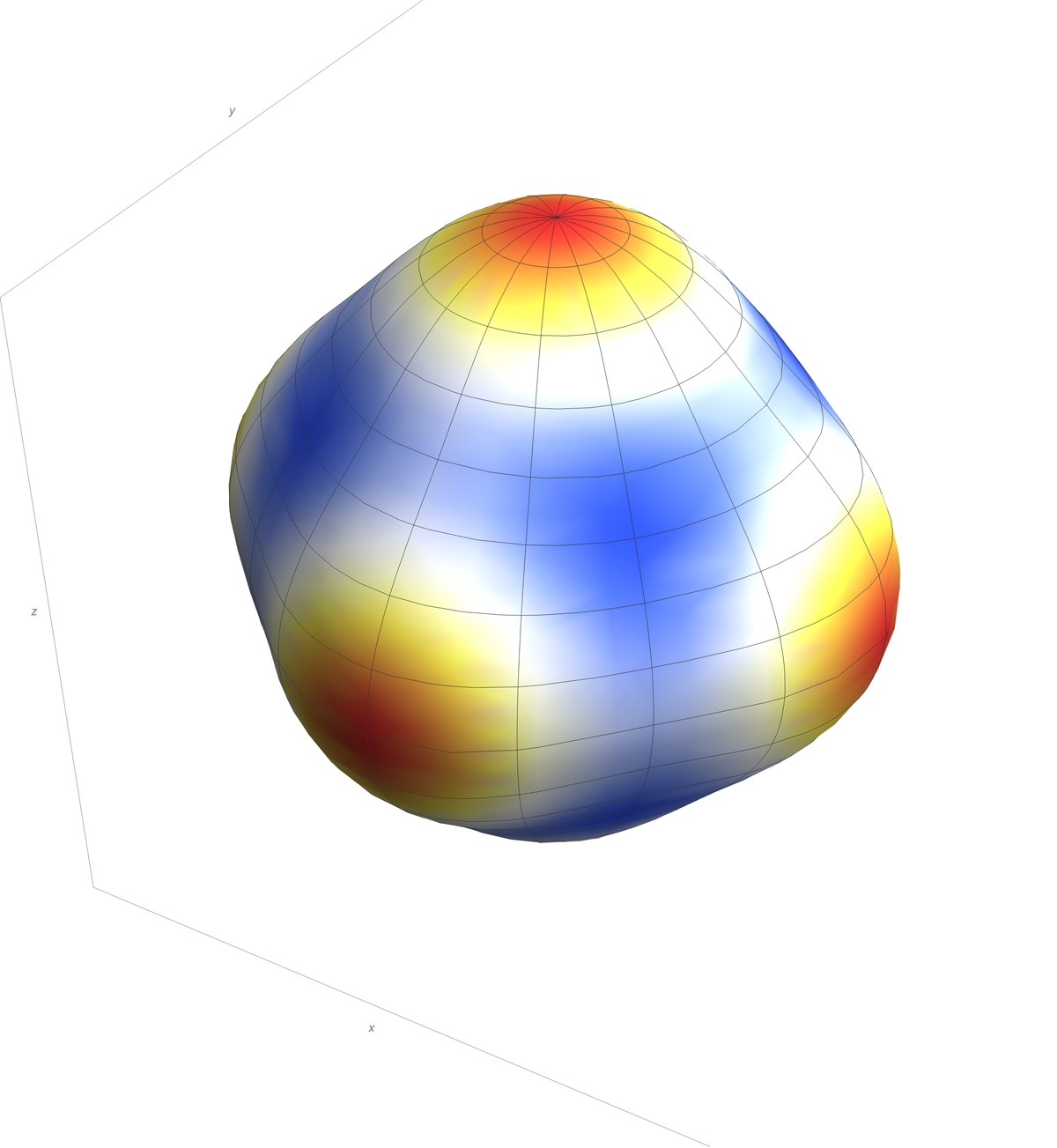}
    \caption{The cosmological horizon for 8 masses located at the vertices of a cube in the bulk plotted in static coordinates. The horizon is shown for an extreme value of $m$, beyond perturbative validity, to highlight the qualitative features of the horizon's shape. The coloring of the horizon is a temperature map with red and blue corresponding to the largest and smallest values of $r'$ respectively. The plot shows that the horizon moves inwards towards the masses leading to an octahedral geometry.}
    \label{fig:cube}
\end{figure}

For our next example, we consider 8 points of mass $m/8$ aligned on the vertices of a cube. The multipole expansion of the Newtonian potential to leading order in $d$ is given by

\begin{equation}
    \label{eq:potential cube}
    \Phi_{cube} = -\frac{m}{r} + \frac{m \left(35 d^4 \left(8 \sin ^4(\theta ) \cos (4 \phi )+4 \cos (2 \theta )+7 \cos (4 \theta )\right)+63 d^4\right)}{128 r^5},
\end{equation}
where $d$ is given in table 1. After the monopole, the leading behavior of the potential is the $L=4$ mode. We match the metric perturbation functions $H_4^M(r)$ to the Newtonian potential and find the constants of integration
\begin{equation}
    \mathbf{c}_4^0 = \frac{14 i \sqrt{\pi } d^4 m}{3 \ell^7} \quad \& \quad \mathbf{c}_4^4 =\mathbf{c}_{4}^{-4} = \frac{i \sqrt{70 \pi } d^4 m}{3 \ell^7},
\end{equation}
with all others being 0. We substitute these constants of integration into the metric perturbation functions to find how the cube affects the spacetime. By following the same procedure outlined above for the dipole, we find the static coordinate location of the horizon to be 
\begin{equation}
    \label{eq: cube location}
    r'_{\mathcal{H}} = \ell - m + d^4 \left(\frac{30 m^3 \sin ^4(\theta ) \cos (4 \phi )}{\ell^6}+\frac{15 m^3 \cos (2 \theta )}{\ell^6}+\frac{105 m^3 \cos (4 \theta )}{4 \ell^6}+\frac{27 m^3}{4 \ell^6}\right).
\end{equation}
The surface given by (\ref{eq: cube location}) is plotted in fig. \ref{fig:cube} in a temperature map and for an  extreme value of $m$ to qualitatively see the behavior of the horizon.

We find that the horizon begins by taking the shape of the dual polyhedron to the cube, the octahedron. As we saw in the case of the dipole, localized matter pulls the horizon inwards towards the vertices of the cube (shown in blue on the horizon). Protrusions of the horizon correspond to faces of the cube, which are shown in red. This interchange of faces and vertices is precisely what defines a dual polyhedron. We confirm that the horizon is indeed an octahedron by finding that it is invariant under the octahedral group. 

We now briefly discuss the surface area of one of the protrusions of the surface. As seen in fig. \ref{fig:cube}, there is a bump at the north pole of the horizon which we find extends from $\theta = 0$ to $\theta_{crit} = \frac{1}{2}\arccos\left(\frac{1}{35}(-5 + 4 \sqrt{30}\right)$ where the horizon location is that of the Schwarzschild de Sitter solution with total mass $m$. The difference between the protruding horizon due to the $L=4$ mode and the $L=0$ mode is \begin{equation}
    \Delta A_{cap} = -\frac{16 \sqrt{\frac{5}{7}-\frac{4 \sqrt{\frac{2}{15}}}{7}} \pi  d^4 m}{35 \ell^3}. 
\end{equation}
Notably, this expression has in it a factor of $d$, the size of the cube in the bulk. In fact, for any shape in the bulk, the corresponding horizon protrusion will have a subregion surface area that encodes the size of that shape. Thus we have shown that all of the information required to determine the Platonic solid deep within the bulk is measurable from the horizon.

\section{Discussion}
We have shown how localized energy configurations deep within the de Sitter bulk affect the cosmological horizon. We primarily considered arrays of masses at the vertices of Platonic solids, but other configurations such as a static mass binary were considered as well. The bulk configurations studied resulted in deformations of the empty de Sitter horizon which inherit the symmetries of the Platonic solid within the bulk. Interestingly, the geometry of the horizon is the dual to the solid within the bulk. Not only do the deformations of the horizon encode the symmetries of the bulk, but they also encode the size of the Platonic solid which can be obtained by studying the peaks and dips around the Schwarzschild de Sitter geometry associated to the total mass of the solid.

While our work only addressed static configurations, future work could extend our analyses to other forms of energy and momentum such as charge and rotation. It remains to be seen how dynamics within the bulk of de Sitter spacetime affect the dynamics of the horizon. Eventually, as the horizon equilibrates, fast scrambling of the horizon's angular dependence will occur.

In fact, one should turn this discussion around and ask the following:\newline Given the data on the cosmological horizon, how can a quantum description defined by the holographic degrees of freedom precisely reproduce these (semi-)classical results? More generally, one could also ask how the quantum theory encodes the dynamics (and not just the statics) in the bulk of spacetime with a positive cosmological constant.

\begin{center} \textbf{Acknowledgements}\end{center}
We thank Phuc Nguyen for early collaboration on this work. We would also like to thank Ted Jacobson and Aaron Zimmerman for discussions regarding the project. The work of WF and SR was supported by the National Science Foundation under grant PHY-2210562. SR thanks the Yukawa Institute at the University of Kyoto for their hospitality during the
YITP-T-23-01 Quantum Information, Quantum Matter and Quantum Gravity workshop where part of this work was completed.

\newpage

\appendix{}

\section{Horizon shapes for other configurations}
We now present the cosmological horizons for configurations of static masses within the de Sitter bulk not present in the main body of the work. We first find the horizons for the rest of the platonic solids. We then discuss a configuration with 2 distinct mass scales.

\subsection{Tetrahedron}
We build a tetrahedron by taking 4 masses, with total mass $m$, arranged on the vertices of a pyramid centered at the origin. The locations of the vertices in Cartesian coordinate of each mass can be found in Table 1. The multipole expansion of the Newtonian potential to leading order in $d$ is given by

\begin{equation}
    \Phi_{tetrahedron} = -\frac{m}{r}+ \frac{5 d^3 m \left(8 \sin ^3(\theta ) \cos (3 \phi )+3 \sqrt{2} \cos (\theta )+5 \sqrt{2} \cos (3 \theta )\right)}{256 \sqrt{3} r^4}.
\end{equation}
We find that the lowest order non-vanishing perturbation after the monopole is the $L=3$ mode. We are only concerned with the dominating angular contribution to the cosmological horizon and apart from the $L=0$ mode only calculate the $L=3$ perturbation. The details for the $L=0$ mode are identical for every configuration of total mass $m$ and are described in the text. The constants of integration of the perturbation functions are found by matching to the above Newtonian potential. The non-zero constants are found to be

\begin{equation}
    \mathbf{c}^{(3,0)} = -\frac{5 i \sqrt{\frac{\pi }{42}} d^3 m}{4 \ell^5} \quad \& \quad \mathbf{c}^{(3,3)} =-\mathbf{c}^{(3,-3)} = \frac{i \sqrt{\frac{5 \pi }{21}} d^3 m}{4 \ell^5}.
\end{equation}

Upon substituting the metric perturbation functions and their constants of integration into the horizon location condition, we find
\begin{equation}
    \label{eq: tet  location}
    \begin{aligned}
    r'_{\mathcal{H}} = \ell - m+ d^3 \Bigg(\frac{5 m^3 \sin ^3(\theta ) \cos (3 \phi )}{2 \sqrt{3} \ell^5}+\frac{25 m^3 \cos ^3(\theta )}{2 \sqrt{6} \ell^5}-\frac{5 \sqrt{\frac{3}{2}} m^3 \cos (\theta )}{2 \ell^5}\Bigg) 
    \end{aligned}
\end{equation}
which is expressed in the same static coordinates as the Schwarzschild de Sitter solution. The correction to the location of the cosmological horizon due to the $L=3$ mode is very small and goes as $\sim d^3 m^3 \sim \ell^2 m^4.$ This leads the horizon to look \textit{approximately} like a sphere. The surface given by (\ref{eq: tet  location}) is plotted below for an extreme value of $m$ to show the qualitative behavior of the horizon. We see dips in the horizon (blue) that correspond to the location of the vertices in the bulk protrusions (red) of the horizon which correspond to faces. The horizon has the geometry of a tetrahedron since the tetrahedron is self-dual. 
\begin{figure}[H]
    \centering
    \includegraphics[width=8cm]{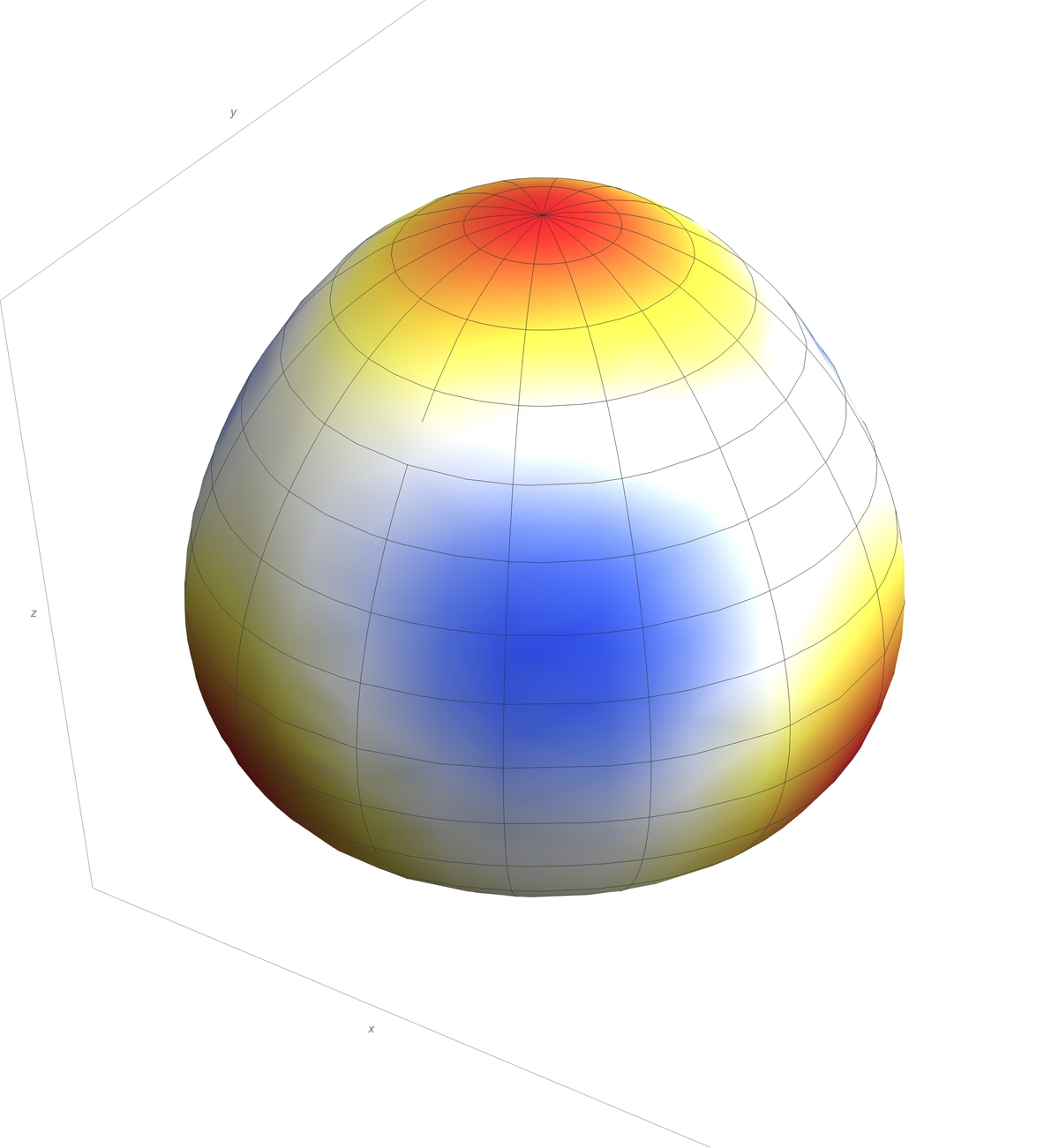}
        \caption{The cosmological horizon for a tetrahedron within the de Sitter bulk is plotted in a temperature map where red corresponds to larger values of $r$ and blue corresponds to smaller values of $r$. We see that since the tetrahedron is self-dual the horizon is also a tetrahedron}
    \label{fig:tet}
\end{figure}

\subsection{Octahedron}
The octahedral configuration is built by taking 6 masses of total mass $m$ and aligning them along the vertices of an octahedron. The locations of the vertices in Cartesian coordinate of each mass can be found in Table 1. The multipole expansion of the Newtonian potential to leading order in $d$ is given by
\begin{equation}
    \Phi_{octahedron} = -\frac{m}{r}-\frac{m \left(35 d^4 \left(8 \sin ^4(\theta ) \cos (4 \phi )+4 \cos (2 \theta )+7 \cos (4 \theta )\right)+63 d^4\right)}{768 r^5}.
\end{equation}
The lowest order non-vanishing perturbation after the monopole is the $L=4$ mode. The constatns of integration for the perturbation function to match to these potentials are 

\begin{equation}
    \mathbf{c}^{(4,0)} = -\frac{7 i \sqrt{\pi } d^4 m}{9 \ell^7} \quad \& \quad \mathbf{c}^{(4,4)} =\mathbf{c}^{(4,-4)} = -\frac{i \sqrt{35 \pi} d^4 m}{9\sqrt{2} \ell^7},
\end{equation}
with all others being 0. Substituting in to the horizon location condition we find
\begin{equation}
    \label{eq: oct  location}
    r'_{\mathcal{H}} = \ell - m + d^4 \left(-\frac{5 m^3 \sin ^4(\theta ) \cos (4 \phi )}{\ell^6}-\frac{5 m^3 \cos (2 \theta )}{2 \ell^6}-\frac{35 m^3 \cos (4 \theta )}{8 \ell^6}-\frac{9 m^3}{8 \ell^6}\right).
\end{equation}
to be the location of the cosmological horizon in Schwarzschild de Sitter static coordinates.
The surface given by (\ref{eq: oct  location}) is plotted below for an extreme value of $m$ to highlight the qualitative features of the horizon. We see that that the horizon takes the shape of the cube, the dual Platonic solid to the octahedron. Correspondingly, we saw in the main body of the text that the cosmological horizon for a cube in the bulk is indeed the octahedron. 

\begin{figure}[H]
    \centering
    \includegraphics[width=8cm]{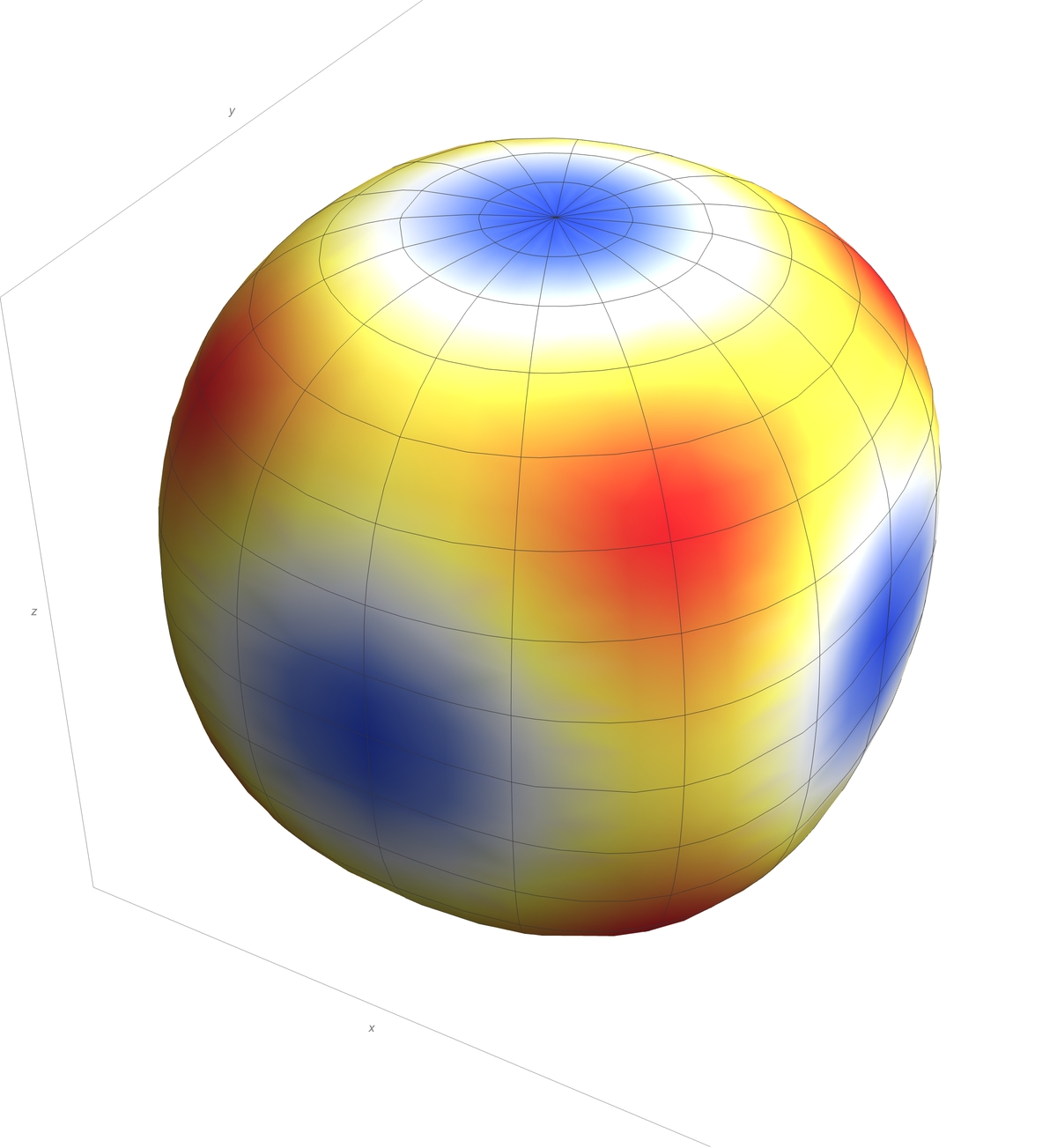}
        \caption{The cosmological horizon for an octahedron within the de Sitter bulk is plotted in a temperature map where red corresponds to larger values of $r$ and blue corresponds to smaller values of $r$. We see that the cosmological horizon becomes a cube, the dual solid to an octahedron.}      \label{fig:dodec}
\end{figure}

\subsection{Icosahedron}
The icosahedron is built from 12 masses with total mass $m$. The locations of the vertices in Cartesian coordinate of each mass can be found in Table 1. The multipole expansion of the Newtonian potential to leading order in $d$ is given by
\begin{equation}
\begin{aligned}
  \Phi_{icosahedron}=-\frac{m}{r} &+  \frac{275 \left(2 \sqrt{5}-5\right) d^6 m}{512 \left(\sqrt{5}-5\right)^6 r^7} \Big(105 \sin (2 \theta +5 \phi )-84 \sin (4 \theta +5 \phi )+21 \sin (6 \theta +5 \phi )\\&+105 \sin (2 \theta -5 \phi )-84 \sin (4 \theta -5 \phi )+21 \sin (6 \theta -5 \phi )+105 \cos (2 \theta )\\&+126 \cos (4 \theta )+231 \cos (6 \theta )+50\Big)
  \end{aligned}
\end{equation}
The lowest order non-vanishing perturbation after the monopole is the $L=6$ mode. The constants of integration to match the perturbation function to the Newtonian potential are found to be
\begin{equation}
    \mathbf{c}^{(6,0)} = \frac{1100 i \left(2 \sqrt{5}-5\right) \sqrt{\frac{\pi }{13}} d^6 m}{\left(\sqrt{5}-5\right)^6 \ell^{11}} \quad \& \quad \mathbf{c}^{(6,5)} =-\mathbf{c}^{(6,-5)} = \frac{100 i \left(2 \sqrt{5}-5\right) \sqrt{\frac{77 \pi }{13}} d^6 m}{\left(\sqrt{5}-5\right)^6 \ell^{11}}
\end{equation}
with all others being zero. We follow the same procedure of substituting the perturbation equations into the horizon location condition to find the corrected cosmological horizon. However, it is not illustrative to give the expression of the location of the horizon due to its complicated angular dependence. Instead, we plot the surface of the horizon for an extreme value of $m$ below to assess its qualitative features. We see that that the horizon takes the shape of the dodecahedron, the dual platonic solid to the icosahedron. The horizon also inherits the symmetry of the bulk configuration, the icosahedral group. 

\begin{figure}[H]
    \centering
    \includegraphics[width=8cm]{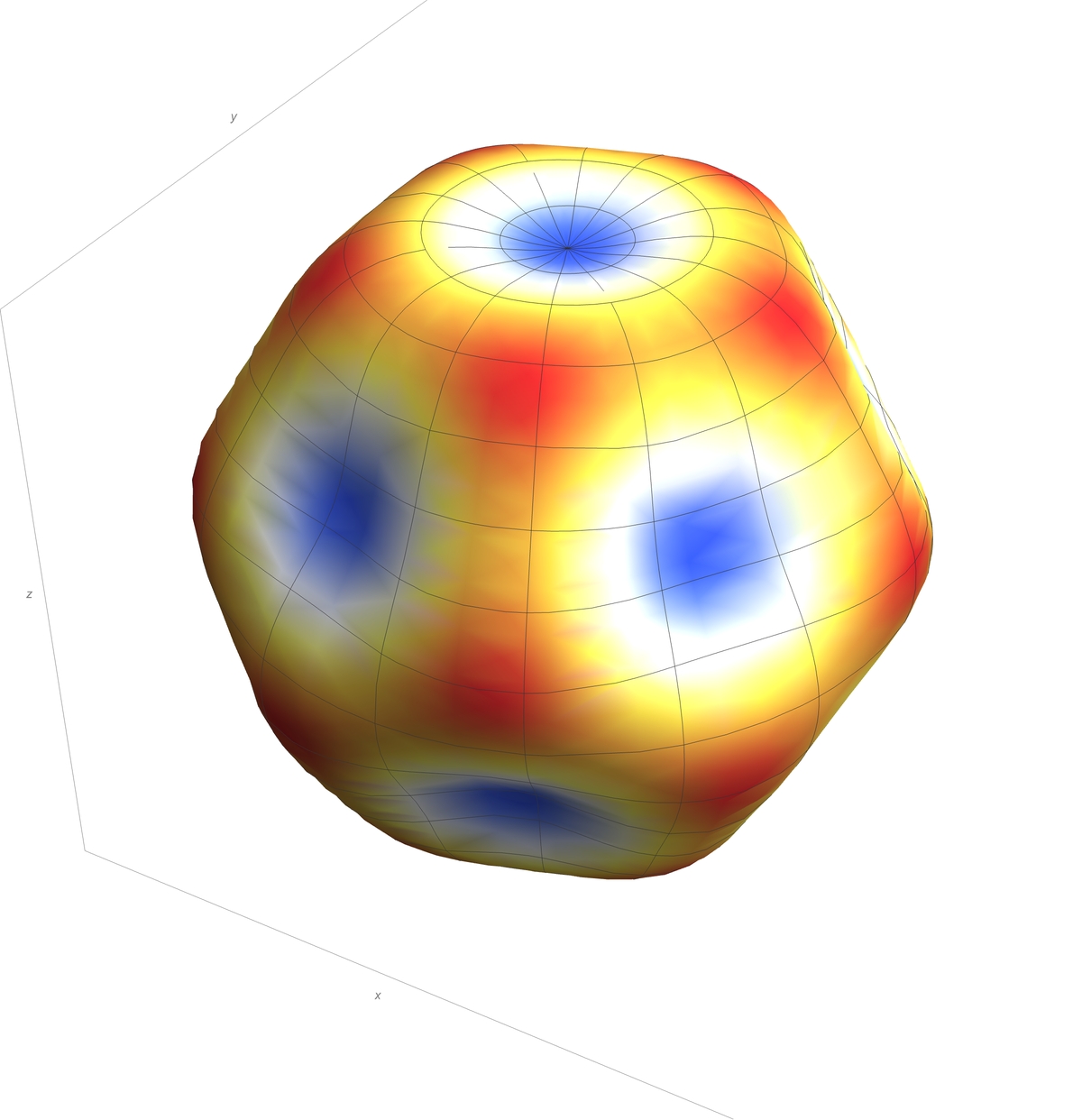}
        \caption{The cosmological horizon for an icosahedron within the de Sitter bulk is plotted in a temperature map where red corresponds to larger values of $r$ and blue corresponds to smaller values of $r$. We see that the cosmological horizon becomes a dodecahedron, the dual solid to an icosahedron.}    \label{fig:ico}
\end{figure}

\subsection{Dodecahedron}
The icosahedron is built from 12 masses with total mass $m$. The locations of the vertices in Cartesian coordinate of each mass can be found in Table 1. The multipole expansion of the Newtonian potential to leading order in $d$ is given by
\begin{equation}
\begin{aligned}
  \Phi_{dodecahedron}=-\frac{m}{r} &+  \frac{33 d^6 m}{128 \left(\sqrt{5}+1\right)^6 r^7} \Bigg(336 \left(9 \sqrt{5}+20\right) \sin ^6(\theta ) \cos (6 \phi )\\&+336 \left(4 \sqrt{5}+9\right) \sin ^4(\theta ) (11 \cos (2 \theta )+9) \cos (4 \phi ))\\&-42 \left(9 \sqrt{5}+20\right) \sin ^2(\theta ) (60 \cos (2 \theta +33 \cos (4 \theta )+35) \cos (2 \phi ))\\&-\left(4 \sqrt{5}+9\right) (105 \cos (2 \theta )+126 \cos (4 \theta )+231 \cos (6 \theta )+50)\Bigg)
  \end{aligned}
\end{equation}
The lowest order non-vanishing perturbation after the monopole is the $L=6$ mode. The constants of integration to match the perturbation function to the Newtonian potential are found to be
\begin{equation}
\begin{aligned}
    \mathbf{c}^{(6,0)} &= -\frac{33 i \sqrt{\frac{\pi }{13}} d^6 m}{4 \ell^{11}}, \quad \mathbf{c}^{(6,2)} = \mathbf{c}^{(6,-2)} = -\frac{264 i \left(9 \sqrt{5}+20\right) \sqrt{\frac{21 \pi }{65}} d^6 m}{\left(\sqrt{5}+1\right)^6 \ell^{11}}\\
   \mathbf{c}^{(6,4)} &= \mathbf{c}^{(6,-4)} = \frac{33 i \sqrt{\frac{7 \pi }{26}} d^6 m}{4 \ell^{11}}, \quad  \mathbf{c}^{(6,6)} = \mathbf{c}^{(6,-6)}  = \frac{24 i \left(9 \sqrt{5}+20\right) \sqrt{\frac{231 \pi }{13}} d^6 m}{\left(\sqrt{5}+1\right)^6 \ell^{11}}
\end{aligned}\end{equation}
with all others being 0. 

We follow the same procedure of substituting the perturbation equations into the horizon location condition to find the corrected cosmological horizon. However, it is not illustrative to give the expression of the location of the horizon due to its complicated angular dependence. Instead, we plot the surface of the horizon for an extreme value of $m$ below to assess its qualitative features. We see that that the horizon takes the shape of the icosahedron, the dual platonic solid to the dodecahedron. The horizon also inherits the symmetry of the bulk configuration, the icosahedral group. 

\begin{figure}[H]
    \centering
    \includegraphics[width=8cm]{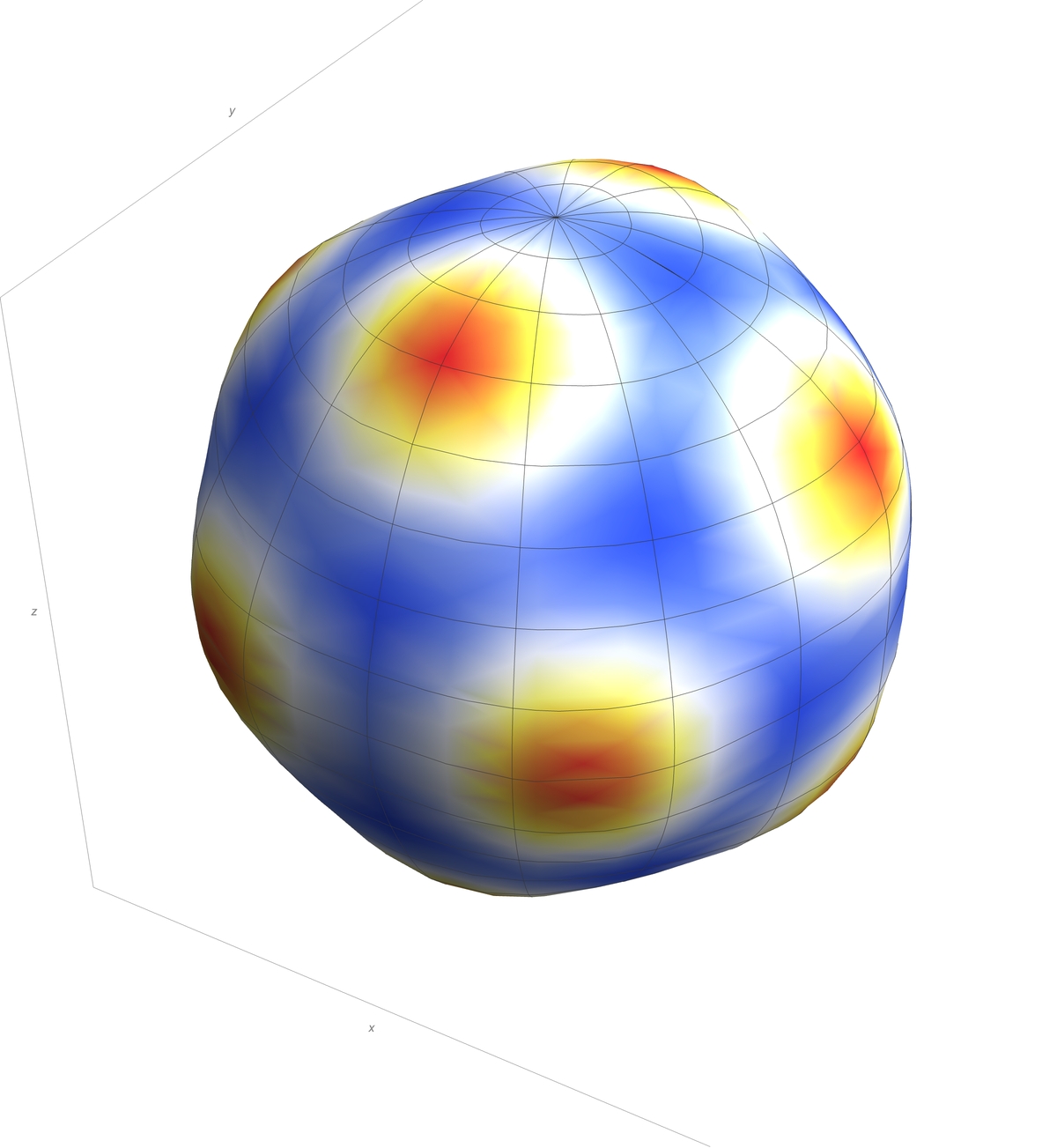}
        \caption{The cosmological horizon for a dodecahedron within the de Sitter bulk is plotted in a temperature map where red corresponds to larger values of $r$ and blue corresponds to smaller values of $r$. We see that the cosmological horizon becomes an icosahedron, the dual solid to a dodecahedron.}      \label{fig:dodec}
\end{figure}

\subsection{Pair of static binaries}
We consider a configuration of masses built by two pairs of static binaries, aligned along the $z$ and $x$ axes with total binary mass $m_z$ and $m_x$ respectively. The Newtonian potential for such a configuration is found by taking the superposition of each binary's potential and is found to be 
\begin{equation}
\begin{aligned}
\Phi_{\text{binary pair}} &= \frac{2 (m_z+m_x)}{r}+\frac{d_z^2 m_z (3 \cos (2 \theta )+1)}{2 r^3}\\&-\frac{d_x^2 m_x \left(\cos ^2(\theta )-\frac{1}{2} \sin ^2(\theta ) (3 \cos (2 \phi )+1)\right)}{r^3}
\end{aligned}
\end{equation}
to leading order in the dipole sizes $d_x$ and $d_z$. 

The location of each mass $\pm d_z \mathbf{e}_z$ and $\pm d_x \mathbf{e}_x$ are found by solving the coupled equations 
\begin{equation}
    \begin{aligned}
            \frac{m_z}{4 dz^2} &= d_z \left(\frac{1}{\ell^2} - \frac{ 2m_x}{(d_x^2+dz^2)^{3/2}}\right)\\
        \frac{m_x}{4 dx^2} &= d_x \left(\frac{1}{\ell^2} - \frac{ 2m_z}{(d_x^2+dz^2)^{3/2}}\right)
    \end{aligned}
\end{equation}
which arise from the static mass condition (\ref{eq:static condition}).

In the metric, we treat each binary pair of masses separately when constructing the corresponding metric perturbations. We then build the solution by taking a superposition of of the two perturbations about the empty de Sitter spacetime background metric. The binary aligned along the $z$-axis is identical to the treatment in section 5, except the mass and distance parameters $m$ and $d$ are replaced by $m_z$ and $d_z$. For the binary aligned along the $x$-axis, we find the non-zero constants of integration of the metric perturbation functions to be

\begin{equation}
    \mathbf{c}^{(2,0)} =\frac{2 i \sqrt{\frac{\pi }{5}} d_x^2 m_x}{\ell^3} \quad \& \quad \mathbf{c}^{(2,2)} =\mathbf{c}^{(2,-2)} = -\frac{i \sqrt{\frac{6 \pi }{5}} d_x^2 m_x}{\ell^3}.
\end{equation}

After combining the two perturbations the horizon location is given by
\begin{equation}
    \label{eq: loz  location}
    \begin{aligned}
    r'_{\mathcal{H}} = \ell - m_z - m_x & -\frac{\ell}{2}  \Bigg(\frac{32 d_z^2 m_z \left(3 \cos ^2(\theta )-1\right) \left(m_x^2+2 m_x m_z+m_z^2\right)}{\ell^5}\\&-\frac{8 d_x^2 m_x (m_x+m_z)^2 \left(-3 \sin ^2(\theta ) \cos (2 \phi )+3 \cos ^2(\theta )-1\right)}{\ell^5}\Bigg).
    \end{aligned}
\end{equation}
We see that the horizon encodes both masses in the spacetime, and the angular dependence of the horizon depends on the interplay of the size of the masses. We plot the horizon location below for extreme values of $m_z$ and $m_x$ outside perturbative validity. We take $m_x = 1.5 \times m_z$ to highlight the interplay between the masses and the stretching of the horizon along both $x$ and $z$. 
\begin{figure}[h]
    \centering
    \includegraphics[width=8cm]{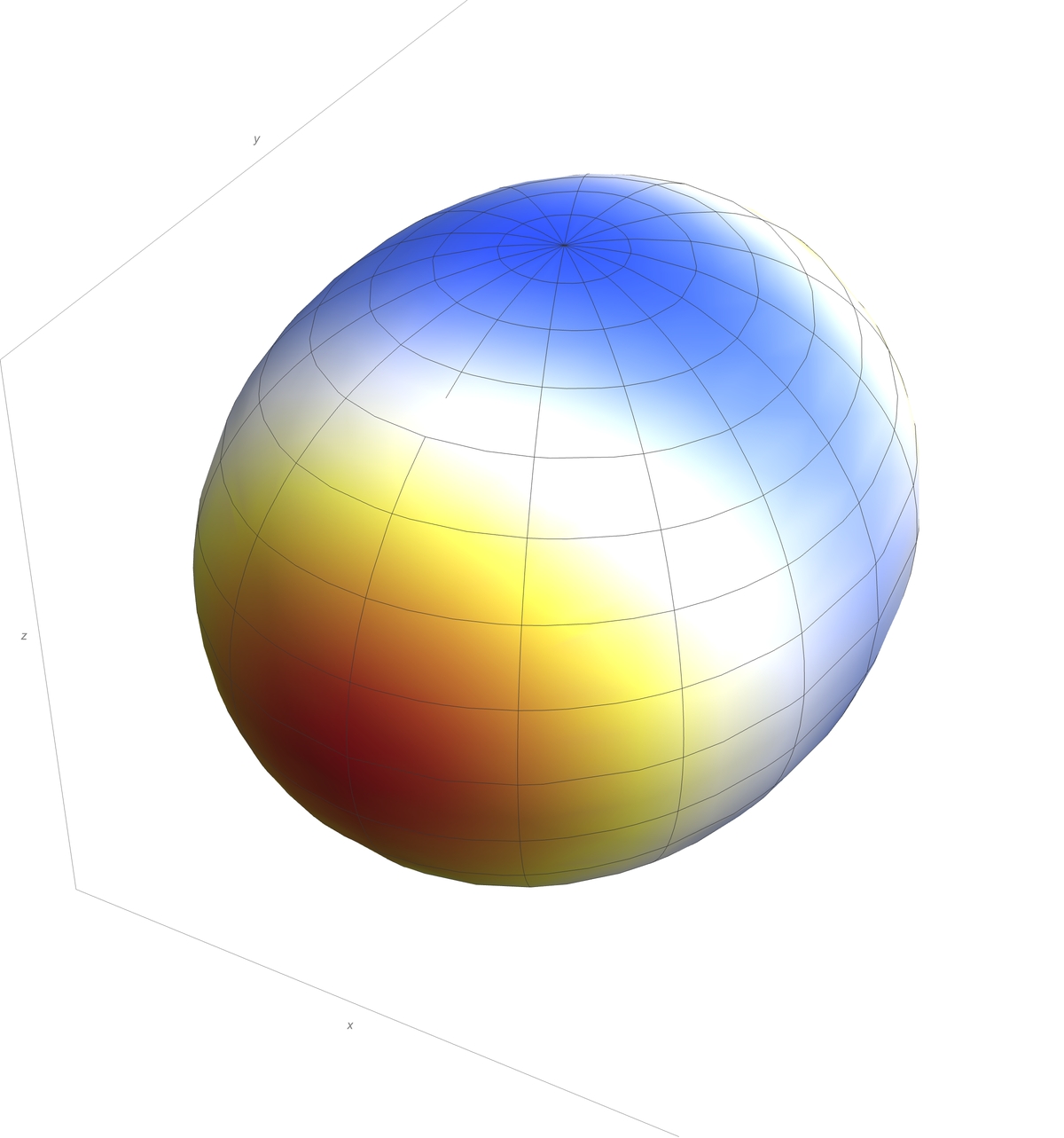}
        \caption{The cosmological horizon for two pairs of static binaries, one aligned along the $x$-axis and one aligned along the $z$-axis, with total mass $m_x$ and $m_z$ respectively. In this figure $m_x = 1.5 \times m_z$. The horizon begins to take a dumbbell shape that bulges most along the $y$ direction.}       \label{fig:lozenge}
\end{figure}
\newpage

\bibliographystyle{JHEP} 
\bibliography{bib} 

\providecommand{\href}[2]{#2}\begingroup\raggedright\begin{thebibliography}{10}

\bibitem{hep-th/9905111}
O.~Aharony, S.S.~Gubser, J.M.~Maldacena, H.~Ooguri and Y.~Oz, \emph{{Large N
  field theories, string theory and gravity}},
  \href{https://doi.org/10.1016/S0370-1573(99)00083-6}{\emph{Phys. Rept.}
  {\bfseries 323} (2000) 183}
  [\href{https://arxiv.org/abs/hep-th/9905111}{{\ttfamily hep-th/9905111}}].

\bibitem{hep-th/0102077}
T.~Banks and W.~Fischler, \emph{{M theory observables for cosmological
  space-times}},  \href{https://arxiv.org/abs/hep-th/0102077}{{\ttfamily
  hep-th/0102077}}.

\bibitem{Fischler2000}
W.~Fischler, \emph{Taking de {{Sitter Seriously}}. {{Talk}} given at {{Role}}
  of {{Scaling Laws}} in {{Physics}} and {{Biology}} ({{Celebrating}} the 60th
  {{Birthday}} of {{Geoffrey West}}), in {{Santa Fe}}},  2000.

\bibitem{Banks2000}
T.~Banks, \emph{Talk at the 60th {{Birthday Celebration}} of {{L. Susskind}},
  at {{Stanford University}}},  2000.

\bibitem{hep-th/0007146}
T.~Banks, \emph{{Cosmological breaking of supersymmetry?}},
  \href{https://doi.org/10.1142/S0217751X01003998}{\emph{Int. J. Mod. Phys. A}
  {\bfseries 16} (2001) 910}
  [\href{https://arxiv.org/abs/hep-th/0007146}{{\ttfamily hep-th/0007146}}].

\bibitem{hep-th/0609062}
T.~Banks, B.~Fiol and A.~Morisse, \emph{{Towards a quantum theory of de Sitter
  space}}, \href{https://doi.org/10.1088/1126-6708/2006/12/004}{\emph{JHEP}
  {\bfseries 12} (2006) 004}
  [\href{https://arxiv.org/abs/hep-th/0609062}{{\ttfamily hep-th/0609062}}].

\bibitem{hep-th/9806039}
W.~Fischler and L.~Susskind, \emph{{Holography and cosmology}},
  \href{https://arxiv.org/abs/hep-th/9806039}{{\ttfamily hep-th/9806039}}.

\bibitem{2306.05264}
S.~A, T.~Banks and W.~Fischler, \emph{{Quantum theory of three-dimensional de
  Sitter space}},
  \href{https://doi.org/10.1103/PhysRevD.109.025011}{\emph{Phys. Rev. D}
  {\bfseries 109} (2024) 025011}
  [\href{https://arxiv.org/abs/2306.05264}{{\ttfamily 2306.05264}}].

\bibitem{2109.01322}
L.~Susskind, \emph{{Black Holes Hint towards De Sitter Matrix Theory}},
  \href{https://doi.org/10.3390/universe9080368}{\emph{Universe} {\bfseries 9}
  (2023) 368} [\href{https://arxiv.org/abs/2109.01322}{{\ttfamily
  2109.01322}}].

\bibitem{2209.09999}
L.~Susskind, \emph{{De Sitter Space, Double-Scaled SYK, and the Separation of
  Scales in the Semiclassical Limit}},
  \href{https://arxiv.org/abs/2209.09999}{{\ttfamily 2209.09999}}.

\bibitem{Gibbons1977}
G.W.~Gibbons and S.W.~Hawking, \emph{Cosmological event horizons,
  thermodynamics, and particle creation},
  \href{https://doi.org/10.1103/PhysRevD.15.2738}{\emph{Physical Review D}
  {\bfseries 15} (1977) 2738}.

\bibitem{2206.10780}
V.~Chandrasekaran, R.~Longo, G.~Penington and E.~Witten, \emph{{An algebra of
  observables for de Sitter space}},
  \href{https://doi.org/10.1007/JHEP02(2023)082}{\emph{JHEP} {\bfseries 02}
  (2023) 082} [\href{https://arxiv.org/abs/2206.10780}{{\ttfamily
  2206.10780}}].

\bibitem{hep-th/0212209}
N.~Goheer, M.~Kleban and L.~Susskind, \emph{{The Trouble with de Sitter
  space}}, \href{https://doi.org/10.1088/1126-6708/2003/07/056}{\emph{JHEP}
  {\bfseries 07} (2003) 056}
  [\href{https://arxiv.org/abs/hep-th/0212209}{{\ttfamily hep-th/0212209}}].

\bibitem{Regge1957}
T.~Regge and J.A.~Wheeler, \emph{Stability of a {{Schwarzschild Singularity}}},
  \href{https://doi.org/10.1103/PhysRev.108.1063}{\emph{Physical Review}
  {\bfseries 108} (1957) 1063}.

\bibitem{Zerilli1970}
F.J.~Zerilli, \emph{Effective {{Potential}} for {{Even-Parity Regge-Wheeler
  Gravitational Perturbation Equations}}},
  \href{https://doi.org/10.1103/PhysRevLett.24.737}{\emph{Physical Review
  Letters} {\bfseries 24} (1970) 737}.

\bibitem{Guven1990}
J.~Guven and D.~N{\'u}{\~n}ez, \emph{Schwarzschild-de {{Sitter}} space and its
  perturbations},
  \href{https://doi.org/10.1103/PhysRevD.42.2577}{\emph{Physical Review D}
  {\bfseries 42} (1990) 2577}.

\bibitem{math-ph/0303071}
M.~Atiyah and P.~Sutcliffe, \emph{{Polyhedra in physics, chemistry and
  geometry}},  \href{https://arxiv.org/abs/math-ph/0303071}{{\ttfamily
  math-ph/0303071}}.

\bibitem{hep-th/0201101}
R.A.~Battye, G.W.~Gibbons and P.M.~Sutcliffe, \emph{{Central configurations in
  three-dimensions}}, \href{https://doi.org/10.1098/rspa.2002.1061}{\emph{Proc.
  Roy. Soc. Lond. A} {\bfseries 459} (2003) 911}
  [\href{https://arxiv.org/abs/hep-th/0201101}{{\ttfamily hep-th/0201101}}].

\bibitem{hep-th/0308200}
G.W.~Gibbons and C.E.~Patricot, \emph{{Newton-Hooke space-times, Hpp waves and
  the cosmological constant}},
  \href{https://doi.org/10.1088/0264-9381/20/23/016}{\emph{Class. Quant. Grav.}
  {\bfseries 20} (2003) 5225}
  [\href{https://arxiv.org/abs/hep-th/0308200}{{\ttfamily hep-th/0308200}}].

\bibitem{2303.07361}
O.J.C.~Dias, G.W.~Gibbons, J.E.~Santos and B.~Way, \emph{{Static Black Binaries
  in de Sitter Space}},
  \href{https://doi.org/10.1103/PhysRevLett.131.131401}{\emph{Phys. Rev. Lett.}
  {\bfseries 131} (2023) 131401}
  [\href{https://arxiv.org/abs/2303.07361}{{\ttfamily 2303.07361}}].

\bibitem{Thomson1904}
J.~Thomson, \emph{{{XXIV}}. {{On}} the structure of the atom: An investigation
  of the stability and periods of oscillation of a number of corpuscles
  arranged at equal intervals around the circumference of a circle; with
  application of the results to the theory of atomic structure},
  \href{https://doi.org/10.1080/14786440409463107}{\emph{The London, Edinburgh,
  and Dublin Philosophical Magazine and Journal of Science} {\bfseries 7}
  (1904) 237}.

\bibitem{Altmann1957}
S.L.~Altmann, \emph{On the symmetries of spherical harmonics},
  \href{https://doi.org/10.1017/S0305004100032370}{\emph{Mathematical
  Proceedings of the Cambridge Philosophical Society} {\bfseries 53} (1957)
  343}.

\bibitem{Gelessus1995}
A.~Gelessus, W.~Thiel and W.~Weber, \emph{Multipoles and {{Symmetry}}},
  \href{https://doi.org/10.1021/ed072p505}{\emph{Journal of Chemical Education}
  {\bfseries 72} (1995) 505}.

\bibitem{arXiv:0707.3222}
T.~Jacobson, \emph{{When is g(tt) g(rr) = -1?}},
  \href{https://doi.org/10.1088/0264-9381/24/22/N02}{\emph{Class. Quant. Grav.}
  {\bfseries 24} (2007) 5717}
  [\href{https://arxiv.org/abs/0707.3222}{{\ttfamily 0707.3222}}].

\end{thebibliography}\endgroup

\end{document}